\documentclass{PoS}
\newcommand{\ber}{\begin{eqnarray}}
\newcommand{\eer}[1]{\label{#1}\end{eqnarray}}
\def\lagr{{\cal L}}
\def\ham{{\cal H}}
\def\half{{\textstyle \frac 1 2}}
\def\+{{+\!\!\!+}}
\def\pp{\mbox{\tiny${}_{\stackrel\+ =}$}}
\newcommand{\re}[1] {(\ref{#1})}
\newcommand{\nll}{N\!=\!(1,1)}
\newcommand{\nn}{\nonumber}
\newcommand{\na}{\nabla}

\newcommand{\pa}{\partial}

\title{Covariant Hamiltonians, sigma models and supersymmetry}

\ShortTitle{Hamiltonian Sigma Model}

\author{\speaker{Ulf Lindstr\"om}\thanks{A footnote may follow.}\\
        Uppsala University, Uppsala and Imperial College, London\\
        E-mail: \email{ulf.lindstrom@physics.uu.se}}

\abstract{We introduce a phase space with spinorial momenta, corresponding to fermionic derivatives, for a $2d$ supersymmetric $(1,1)$ sigma model. We show that there is a generalisation of the covariant De Donder-Weyl Hamiltonian formulation on this phase space with canonical equations equivalent to the Lagrangian formulation, find the corresponding multisymplectic form and Hamiltonian multivectors. The covariance of the formulation makes it possible to see how additional non-manifest supersymmetries arise in analogy to those of the Lagrangian formulation.

We then observe that an intermediate phase space Lagrangian defined on the sum of the tangent and cotanget spaces  is a first order Lagrangian for the sigma model and derive additional supersymmetries for this.}

\FullConference{Corfu Summer Institute 2019 "School and Workshops on Elementary Particle Physics and Gravity" (CORFU2019)\\
		31 August - 25 September 2019\\
		Corfu, Greece}

\begin{document}

\section{Introduction}
In \cite{Lindstrom:2004eh} and \cite{Lindstrom:2004iw}  generalised sigma models with additional auxiliary coordinates resembling spinorial momenta were investigated. Additional supersymmetries were expected to correspond to Generalised Geometry. In particular it was hoped that the Gualtieri map from bihermitean to Generalised K\"ahler geometry, known from \cite{Gualtieri:2003dx} would emerge. The analysis in the Lagrangian formalism was difficult and only partial results, such as the relation for $(1,0)$ models, were obtained. The explicit map from sigma models to Generalised K\"ahler Geometry, yielding the Gualtieri map as a necessesary and sufficient condition for $(2,2)$ supersymmetry, was first derived in \cite{Bredthauer:2006hf} using a Hamiltonian formulation. It may therefore be of some interest to further reformulate the $(1,1)$ sigma model in a covariant Hamiltonian form which lends itself to interpretation in a generalised tangent space. The third  section of this contribution contains a starting point for such a formulation. The Legendre transformation involved in such a formulation naturally suggests a first order formulation of the theory where additional supersymmetries can be studied. The fourth section reports on the results of this study.
\section{The sigma model Lagrangian}
We shall be interested in
the sigma model 
\ber
\int d^2\xi d\theta^+ d\theta^-\lagr = \int_\Sigma d^2\xi d\theta^+ d\theta^-( D_+\phi^iE_{ij}(\phi)D_-\phi^j)~,
\eer{act1}
where
\ber
E_{ij}=(E_{(ij)}+E_{[ij]}):=G_{ij}+B_{ij}~.
\eer{}
The coordinates on the $2d$ superspace domain $\Sigma$  are the bosonic light cone coordinates
$(\xi^\+,\xi^=)$ and the spinorial coordinates $(\theta^+,\theta^-)$, while
$\phi=\phi(\xi^\+,\xi^=,\theta^+,\theta^-)$ are superfields and the spinorial and bosonic derivatives $D_\pm$ and $\partial_{\pp}$ obey
\ber
D^2_\pm=i\partial_{\pp}~.
\eer{}

The field equations that result from varying the action \re{act1} are
\ber
\nabla_+^{(+)}D_-\phi^i=0~,
\eer{feq}
where the the Levi-Civita connection $\Gamma^{(0)}$ has been augmented by torsion\footnote{$G^{ij}G_{jk}=\delta^i_k$}
\ber
{\Gamma^{(\pm)}}_{ij}^{~~k}={\Gamma^{(0)}}_{ij}^{~~k}\pm\half H_{ijm}G^{mk}~,
\eer{gplus}
and  
\ber
H_{ijm}:=B_{ij,m}+cycl. 
\eer{hdef}

\section{The sigma model De Donder-Weyl Hamiltonian}
For bosonic field theories there is  an alternative formulation to Lagrangian field theories where one introduces a momentum dual to each derivative of the field, spacelike as well as timelike. A De Donder-Weyl Hamiltonian  $H_{DW}$ is introduced via  Legendre transforms and the evolution is then given by its canonical equations, the De Donder-Weyl equations \cite{dD} \cite{HW}-\cite{Forger:2002ak}. We now modify and apply those ideas to $\nll$~ superspace.

It was observed in \cite{Lindstrom:2004eh} that a first order action for \re{act1} can be found if we define the spinorial ``momenta'' $S^\pm$ {}\footnote{The functional derivative is taken to act from the left. The first order action will appear in Sec.4.}
\ber
S_{\pm i}=\frac{\partial \lagr}{\partial D_\pm \phi ^i}~.
\eer{mom}
We find
\ber\nn
&&S^+_i=E_{ij}D_-\phi^j\\[1mm]
&&S^-_i=-D_+\phi^jE_{ji}~.
\eer{Sconstr}
{From  \re{Sconstr}, it follows that $\na_+S_{-i}-\na_-S_{+i}\sim \na (B\na\phi)$, and hence $\na_+S^+_i=-\na_-S^-_i$ when $ B=0$, a result sometimes needed in what follows.}

Letting $\alpha:=(+,-)$, a  Legendre transformation $D\phi_\pm \to S^\pm$ is given by
\ber
S^{\alpha}_iD_{\alpha}\phi^i+\lagr~,
\eer{pslag}
together with \re{mom} and yields
\ber
\ham_{DW}=S^{-}_iE^{ij}S^{+}_j~,
\eer{DWe}
with $E^{ij}E_{jk}=\delta^i_k$. Notice that, unlike the usual Hamiltonian, $\ham_{DW}$ is still fully $(1,1)$ superpoincar\'e covariant.

The above formulation represents a model on the sum  two copies of the cotangent space
$\mathbb{T}^ *\oplus\mathbb{T}^ *$. If we extend it by including a copy of the Lagrangian in \re{act1} we  have a model on $\mathbb{T}\oplus\mathbb{T}^ *\oplus\mathbb{T}^ *$.
which may be used to study generalised geometry. In Sec.\ref{addsusy} below we find its extended supersymmetries.  Before turning to these, however, it is worth making a few more comments on the covariant formalism, leaving the details for a future publication \cite{progress}.

\subsection{The equivalence}
In analogy to the usual Canonical equations for a Hamiltonian, we consider the following
\ber\nn
&&D_\alpha\phi^i=\frac {\partial \ham_{DW}}{\partial S^\alpha_i}\\[1mm]
&&D_\alpha S^\alpha_i=\frac {\partial \ham_{DW}}{\partial \phi^i}~.
\eer{can}
We shall call this set of equations the De Donder-Weyl equations for the $\nll$~sigma model.
The two first equations yield the expressions \re{mom} for the momenta. When inserted in the third equation, the field equation
\re{feq} is recovered. In other words, the system \re{can} is an equivalent formulation of the evolution.

\subsection{A multisymplectic form and  Hamiltonian multivector}
In this and the following subsection, we closely follow and adapt the bosonic case as described in \cite{Kanatchikov:1993rp}.

The bosonic DW equations corresponding to \re{DWe} can are related to the existence of a multisymplectic form $\Omega$ and a Hamiltonian multivector field $X$ such that
\ber
X\mbox{\Large $\lrcorner $}~\Omega=d\ham_{DW}
\eer{contr}
For a $n$  dimensional underlying manifold, i.e., for the case of $n$  momenta, and with $\ham_{DM}$ a scalar, $\Omega$ is an $n+1$ form and $X$ a $n-$vector. 

For the case of two fermionic momenta, we look for an analogous formula.The two-vector is
\ber
X=X^{{M_1}{M_2}}\pa_{M_1}\wedge\pa_{M_2} \to {X^{{A}{\alpha}}\pa_{A}\wedge E_{\alpha}} ~,
\eer{x}
where $M$ runs over $(A,\alpha)$, the antisymmetrisation is graded and the index $A=({}^\alpha_i, {}^j)$ corresponding to $S^\alpha_i$ and $\phi^j$ and $E_M=(\pa_A~,E_\alpha)$.  The multisympletic form is
\ber
\Omega:=C_{\alpha\beta}E^\beta\wedge d\phi^i\wedge dS^\alpha_i~.
\eer{o}
Using \re{x} and \re{o} in \re{contr} results in the relations
\ber\nn
&& X^i_\alpha=\frac {\partial \ham_DW}{\partial S^\alpha_i}\\[1mm]
&&X_{i\beta}^{~~\beta}=\frac {\partial \ham_DW}{\partial \phi^i}~.
\eer{}
The solution
\ber\nn
&&X^i_\alpha=D_\alpha \phi^i\\[1mm]
&&X_{i\beta}^{~~\beta}=D_\beta S_i^{~\beta}~
\eer{}
reproduces the De Donder-Weyl equations \re{can}.
{\subsection{A generalised Poisson bracket and conjugate momenta}
Given the multisymplectic form $\Omega$, in the purely even case it is possible to relate it to a generalised Poisson bracket  $\{~,~\}_{GP}$. 
For $n-1$ forms $F=F^\mu dx_\mu$ with Hamiltonian multivector $X_F$ 
\ber
X_F\mbox{\Large $\lrcorner $}~\Omega=dF
\eer{XF}
the bracket with $H$ is
\ber
\{F,H\}_{GP}=(-)^{n-1}X_F\mbox{\Large $\lrcorner $}~dH
\eer{}
Using the DW equations, this means that 
\ber
\star^{-1}dF=\{F,H\}_{GP}~,
\eer{GP}
on the motion. Here $\star$ is the Hodge dual. It follows that $\{F,H\}_{GP}=0$ for conserved quantities.

In the present superspace  case we note that  \re{GP} leads to the one forms $Q^i_\alpha:=\phi^iE_\alpha$ and $S_i:=S_i^\alpha E_\alpha$ satisfying
\ber\nn
&&\star^{-1}dQ^i_\alpha=\{Q^i_\alpha,H\}_{GP}=\pa^i_\alpha \ham_{DW}~,\\[1mm]
&&\star^{-1}dS_i=\{S_i,H\}_{GP}=\pa_i \ham_{DW}~.
\eer{}
It  also leads to $\phi^i$ and $S_i$ being conjugate quantities
\ber
&&\{\phi^i,S_j\}_{GP}=\delta^i_j~.
\eer{}
Here we leave the brief introduction of a covariant supersymmetric Hamiltonian theory, except for a comment at the end of Sec.\ref{addsusy}, and return to the question of symmetries.
\section{Additional supersymmetries}
\label{addsusy}
The action \re{act1} has additional non-manifest supersymmetries
\ber
\delta \phi ^i = \epsilon^+J^i_{(+)j}D_+\phi^j+\epsilon^-J^i_{(-)j}D_-\phi^j
\eer{susy}
provided that $J_{(\pm)}$ are complex structures that preserve the metric (hermiticity) 
\ber
J^t_{(\pm)}GJ_{(\pm)}=G
\eer{}
and 
\ber
\nabla_i^{(\pm)}J_{(\pm)}=0~,
\eer{}
with connections defined   in \re{gplus} and \re{hdef} \cite{Gates:1984nk}.

We note the simple fact that the transformations \re{susy} translate into transformations for the momenta $S^{\pm}_i$ using \re{susy} and 
the relations \re{Sconstr}.
For the plus-supersymmetry with parameter 
$\epsilon^+$ this gives
\ber\nn
&&\delta S^+_i=E_{ik}\na_-^{(+)}\delta\phi^k+E^{js}S^+_s\left(E_{ij},_k-E_{il}\Gamma^{(+)l}_{jk}\right)\delta\phi^k\\[1mm]
&&\delta S^-_i=-\na_+^{(+)}\delta\phi^kE_{ki}+S^-_sE^{sj}\left(E_{ji},_k-\Gamma^{(+)l}_{jk}E_{li}\right)\delta\phi^k
\eer{gsym}
where
\ber
\delta\phi^k=-\epsilon^+J^k_{(+)l}E^{lm}S^-_m~.
\eer{dels}

When $B=0$, \re{gsym} and \re{dels} imply
\ber\nn
&&\delta S^+_i=\epsilon^+\left(J_i^{~k}\na_-S^-_k-J^{ks}S^-_sS^+_n\Gamma_{ki}^{~~n}\right)\\[1mm]
&&\delta S^-_i=-\epsilon^+\left(J_i^{~k}\na_+S^-_k+J^{ks}S^-_sS^-_n\Gamma_{ki}^{~~n}\right)\
\eer{nobehold}
These transformations leave the action \re{act1} invariant and close to a supersymmetry algebra. The results { generalize to $B\ne 0$ } and inclusion of the minus transformations.

\section{A first order system}
We notice that there is an intermediate ``phase space Lagrangian'' on $\mathbb{T}\oplus\mathbb{T}^ *$: 
\ber
-S^\alpha_iD_\alpha\phi^i+S^{-}_iE^{ij}(\phi)S^+_{j}~,
\eer{PLag}
giving the first order (parent) action for \re{act1} derived in  \cite{Lindstrom:2004eh}.  Alternatively we think of it as the Legendre transformation inverse to \re{pslag}. To this we may add any amount $\mu$ of the Lagrangian in \re{act1}. Consider
\ber
\mu D_+\phi^iE_{ij}D_-\phi^j-S^\alpha_iD_\alpha\phi^i+S^{-}_iE^{ij}(\phi)S^+_{j}=:{\mathbb{Z}}^t\mathbb{E}\mathbb{Z}~,
\eer{1sto}
where
\ber
{\mathbb{Z}}^t=(D_+\phi^i,D_-\phi^i,S^{-}_i,S^+_{i})
\eer{}
and
\ber
\mathbb{E}:=\left(\begin{array}{cccc}0&\mu E_{ij}&0&0\\
0&0&0&0\\
0&-\delta^i_j&0&-E^{ij}\\
-\delta^i_j&0&0&0\end{array}\right)
\eer{}

This is the kind of action which was investigated in \cite{Lindstrom:2004eh}
 and \cite{Lindstrom:2004iw}
 for additional supersymmetries.  Here we note that the $\mu$ term is invariant under the variations $\delta \phi^i$ in \re{susy}. Considering the $\epsilon^+$ transformations only, the variations of $S$ given in \re{gsym} with this $\delta \phi^i$, make the remaining terms invariant.

\section*{Acknowledgements}
I am grateful for the hospitality of the theory group at Imperial College, London,  support from the EPSRC programme grant ''New
Geometric Structures from String Theory'' EP/K034456/1, support from  Lars Hierta's foundation, as well 
as for discussions with C.~M.~Hull and M.~Ro\v cek.


\begin{thebibliography}{6666}
\bibitem{Lindstrom:2004eh}
  U.~Lindstr\"om,
  ``Generalized N = (2,2) supersymmetric nonlinear sigma models,''
  Phys.\ Lett.\ B {\bf 587} (2004) 216
  doi:10.1016/j.physletb.2004.03.014
  [hep-th/0401100].
  
\bibitem{Lindstrom:2004iw}
   U.~Lindstr\"om, R.~Minasian, A.~Tomasiello and M.~Zabzine,
  ``Generalized complex manifolds and supersymmetry,''
  Commun.\ Math.\ Phys.\  {\bf 257} (2005) 235
  doi:10.1007/s00220-004-1265-6
  [hep-th/0405085].
  
  \bibitem{Gualtieri:2003dx}
M.~Gualtieri, ``Generalized complex geometry'',
math/0401221 [math-dg].

\bibitem{Bredthauer:2006hf}
  A.~Bredthauer,  U.~Lindstr\"om,, J.~Persson and M.~Zabzine,
``Generalized Kahler geometry from supersymmetric sigma models,''
  Lett.\ Math.\ Phys.\  {\bf 77} (2006) 291
  doi:10.1007/s11005-006-0099-x
  [hep-th/0603130].

\bibitem{dD}
Th.~De Donder, 
``Theorie Invariantive du Calcul des Variations''
Nuov. 'ed. (Gauthier-Villars, Paris 1935)

\bibitem{HW}
H. ~Weyl, 
Ann. Math. 36 (1935) 607.

  
\bibitem{Kanatchikov:1993rp}
  I.~V.~Kanatchikov,
  ``On the canonical structure of De Donder-Weyl covariant Hamiltonian formulation of field theory. 1. Graded Poisson brackets and equations of motion,''
  hep-th/9312162.
  
\bibitem{Gotay:1997eg} 
  M.~J.~Gotay, J.~Isenberg and J.~E.~Marsden,
  ``Momentum maps and classical relativistic fields. Part 1: Covariant Field Theory,''
  physics/9801019 [math-ph].
  
\bibitem{Paufler:2001qka} 
  C.~Paufler and H.~Romer,
  ``Geometry of Hamiltonian N vectors in multisymplectic field theory,''
  J.\ Geom.\ Phys.\  {\bf 44}, 52 (2002)
  doi:10.1016/S0393-0440(02)00031-1
  [math-ph/0102008].
  
\bibitem{Forger:2002ak} 
  M.~Forger, C.~Paufler and H.~Roemer,
  ``The Poisson bracket for Poisson forms in multisymplectic field theory,''
  Rev.\ Math.\ Phys.\  {\bf 15}, 705 (2003)
  doi:10.1142/S0129055X03001734
  [math-ph/0202043].


  
\bibitem{Gates:1984nk}
  S.~J.~Gates, Jr., C.~M.~Hull and M.~Ro\v cek,
  ``Twisted Multiplets and New Supersymmetric Nonlinear Sigma Models,''
  Nucl.\ Phys.\ B {\bf 248} (1984) 157.
  
  \bibitem{progress}
  C.~M.~Hull, U.~Lindstr\"om  and M.~Ro\v cek, in preparation.
\end{thebibliography}
\end{document}